\documentstyle[psfig,amssymb]{aipproc2}
\pssilent
\def\met{\mbox{${\hbox{$E$\kern-0.6em\lower-.1ex\hbox{/}}}_T$}} 

\def\etal{{\it et al.\/}}
\title{\large\bf Search for R-parity Violating SUSY in Run~2 at D\O\ }
\author{S.~Banerjee, N.K.~Mondal, V.S.~Narasimham, and N.~Parua\\
Tata Institute of Fundamental Research, Bombay, India}
\date{\normalsize\rm March 25, 1999}
\begin{document}
\maketitle
\begin{abstract}
\\
We present a study of sensitivity of the R-parity violating SUSY searches with the upgraded D\O\ detector in Run II of the Fermilab Tevatron, within a SUGRA framework. We considered the lightest neutralino as an LSP that decays into a lepton and slepton (R-parity violating decay), resulting in $2e + \ge 4$~jets or $2\mu + \ge 4$~jets final state. The analysis, based on scaling of the Run I results, shows that squarks and gluinos with masses up to about 0.6~TeV could be probed with 2~fb$^{-1}$ of Run II data. This work has been done in the context of the BTMSSM Working Group of the Run II SUSY/Higgs Workshop at Fermilab.
\end{abstract}

\section{Physics motivation}

Recent interest in R-parity violating (RPV) SUSY decay modes is motivated by the possible high-$Q^2$ event excess at 
HERA~\cite{HERA}. When interpretation of the excess through first-generation leptoquarks was excluded by the D\O~\cite{D0-lq} 
and CDF~\cite{CDF-lq} experiments, it was suggested~\cite{HERA-RPV} that such an effect could be explained via the $s$-channel 
production of a charm or top squark decaying into the $e + jet$ final state. Both the production and the decay vertices 
would thereby violate R-parity. Although more recent  data has not confirmed the previous event excess, and  despite the combined 
analysis showed that the anomalous events reported by the H1 and ZEUS experiments were unlikely to originate from the production 
of a single $s$-channel narrow resonance~\cite{Bassler}, interest in RPV signatures has not abatted.

The CDF and D\O\ Collaborations have recently performed searches for RPV SUSY~\cite{CDF-RPV,D0-RPV}, and have set new mass limits 
on the RPV SUSY particles. Both experiments focussed their searches on the $\lambda'$ couplings, as motivated by the 
high-$Q^2$ HERA event excess. The results of the D\O\ searches are extended to the 
Run~2 case and the expected sensitivity to the RPV couplings is discussed.

\section{D\O\ Search for RPV neutralino decays}

\begin{figure}[thb]
\centerline{\protect\psfig{figure=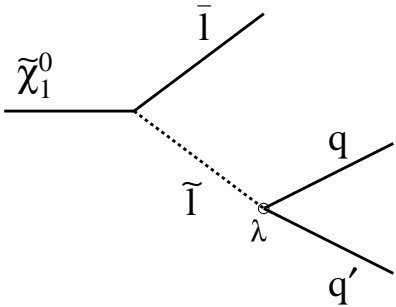,height=1.8in}}
\bigskip
\caption{RPV decay of a neutralino LSP into a lepton and two quarks.}
\label{fig:RPV-decay}
\end{figure}

The D\O\ search for RPV SUSY considered the case of neutralino LSP which decays into a lepton and two quarks due to a finite 
RPV $\lambda'$ coupling (see Fig.~\ref{fig:RPV-decay}). Both the 
electron and muon decay channels were considered, corresponding to what commonly referred to as $\lambda'_{1ij}$ and 
$\lambda'_{2ij}$ couplings, respectively. The corresponding final states contain either $2e  $ or $2\mu $
and at least four accompanying jets. Unlike  at HERA, this search 
is not sensitive to the  value of the RPV coupling, as long as it is large enough so that the neutralino decays within the 
D\O\ detector. That corresponds to $\lambda' \geq 10^{-3}$, which gives a lot of room, given current indirect 
constraints~\cite{Herbi}.

We assume that the neutralino (LSP) pairs are produced in cascade decays of other supersymmetric particles and use all SUSY pair 
production mechanisms when generating signal events.

Signal events were generated within the SUGRA framework with the following values of SUSY parameters: $A_0 = 0$, $\mu < 0$ 
and $\tan\beta = 2$ (the results are not sensitive to the value of $A_0$ .) Center of mass energy of the colliding beams 
was taken to be 2 TeV. {\footnotesize ISAJET}~\cite{ISAJET} was used for event generation. The acceptance and resolution 
of the D\O\ detector were parametrized using the following resolutions:  
$\delta E/E = 2 \% \oplus 15\%/\sqrt{E}~\mbox{[GeV]}$ (electrons), $\delta (1/p)/(1/p) = 0.018 \oplus 0.008 (1/p)$ (muons), and 
$\delta E/E = 3\% \oplus 80\%/\sqrt{E}~\mbox{[GeV]}$ (jets) and found consistent with the full  
detector simulation based on {\footnotesize GEANT}~\cite{GEANT}.

\begin{figure}[htb]
\vspace*{-0.3in}
\centerline{\protect\psfig{figure=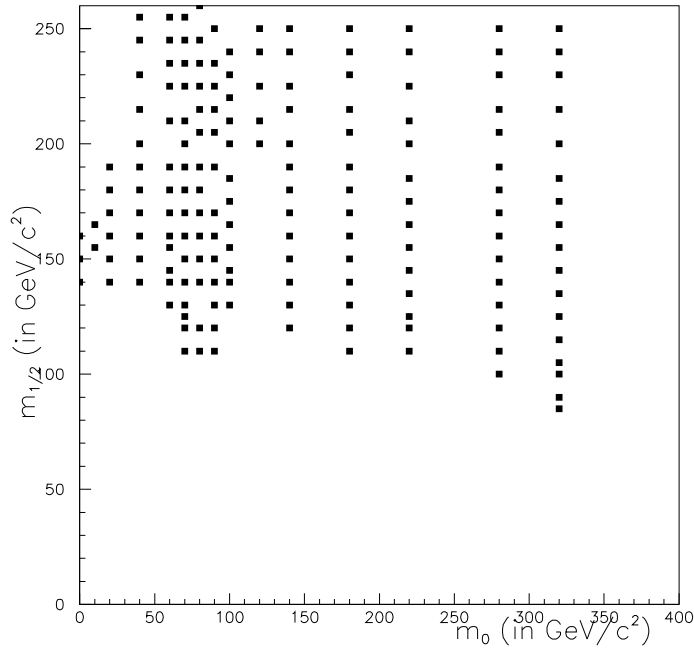,width=4in}}
\caption{Points  in the $(m_0,m_{1/2})$ SUGRA parameter space used to generate RPV events in the $ee + 4$~jets channel.}
\label{fig:RPV-points}
\end{figure}

Figure~\ref{fig:RPV-points} shows the points in the $(m_0,m_{1/2})$ SUGRA parameter space where signal Monte Carlo events 
were generated for the electron channel. Similar points were studied for the muon-decay channel.

\section{Selection criteria for the dielectron channel}

A multijet trigger was used for the analysis of Run~1 data. It was found to be nearly 100\% efficient for the typical RPV signal. 
Since Run~2 trigger list will include a similar trigger, we assume trigger efficiency of 100\% and do not perform any trigger 
simulations for the Run~2.

The following offline selections  were used: 
\begin{itemize}
\item 
At least two good electrons, the leading one with $E_T(e) > 15$ GeV and the other one with $E_T(e) > 10$ GeV;
\item 
Rapidity range $\mid \eta \mid \leq 1.1$ (central calorimeter), or $1.5 \leq \mid \eta \mid \leq 2.5$ (end calorimeters) for all 
the electrons;
\item
Energy isolation for the electrons: the EM energy in the R=0.2 cone about the center of gravity of the EM cluster, subtracted from
the total energy in R=0.4 cone, should not exceed 15\% of the EM energy in the R-0.2 cone.
\item
At least four jets with $E_T(j) > 15 $ GeV and $\mid \eta \mid < 2.5$;
\item
The dielectron invariant mass ($M_{ee}$) should not be in the Z-mass interval, ie, $ \mid M_{ee} - M_Z \mid > 15$ GeV/$c^2$.
\end{itemize}

In the present analysis we have dropped the requirement on $H_T = \sum E_T(e) + \sum E_T(j)$ , but retained all other offline 
criteria that were used in the previous analysis of data from 
Run~I~\cite{D0-RPV}.

\section{Selection in the dimuon channel}

The following event selection requirements were used for the muon decay channel:
\begin{itemize}
\item Two muons, the leading one with $p_T >$ 15 GeV, and the other one with $p_T >$ 10 GeV.
\item Rapidity range $|\eta| < 2.3$ for both muons.
\item Energy isolation requirement for both muons, i.e. the calorimeter energy accompanying the muon in a ($\eta$ $\phi$) cone of 
0.4 should be consistent with that from a minimum ionising particle.
\item At least four jets with $E_T(j) > 15$ GeV and $|\eta| < 2.5$;
\end{itemize}

\section{Signal efficiencies} 
The number of signal events expected can be written as: $\langle N \rangle = \mathcal{L} \cdot \sigma \cdot \epsilon$, where 
$\langle N \rangle$ is the expected number of events for luminosity $ \mathcal{L} $, $\sigma$ is the cross-section, and 
$\epsilon$ is the overall efficiency. The efficiency $\epsilon$ can be split into three terms: 
$\epsilon = {\epsilon}_{\rm trig} \cdot {\epsilon}_{\rm kin} \cdot {\epsilon}_{\rm id}$. Here $\epsilon_{\rm trig}$ is the 
trigger efficiency for the events that pass the offline cuts ( assumed to be 100\%), $\epsilon_{\rm kin}$ is the 
efficiency for offline criteria, which includes kinematic, fiducial and topological requirements, and 
$\epsilon_{\rm id}$ is the electron/jet 
identification efficiency.

The efficiency for identifying jets is very high ($> 95\%$) and is expected to stay the same in Run 2.

Electron identification efficiencies in Run 1 were $80 \pm 7\%$ in the central ($|\eta| < 1.1$) and $71 \pm 7\%$ in the forward 
($1.5 < |\eta| < 2.5$) regions~\cite{D0-RPV}. These efficiencies were calculated for electrons with $E_T(e) > 25$~GeV, It drops 
by about 30\% for electrons with $E_T(e) = 10$~GeV. 

The muon identification efficiencies used in Run 1 were $62 \pm 2\%$ in the central ($|\eta| < 1.0$) and $24 \pm 4\%$ in the 
forward ($1.0 < |\eta| < 1.7$) regions~\cite{rvmu2}. These were calculated for muons with $p_T > 15$~GeV. For muons with 
$10~\mbox{GeV} < p_T < 15$~GeV the efficiencies were 80\% smaller on average~\cite{rvmu3}.

In the present analysis we have taken the overall particle identification efficiency to be $0.90 \pm 0.09$ in each channel, 
independent 
of lepton $E_T$, primarily due to the expectation of a better tracker and muon spectrometer for the upgraded D\O\ experiment.

\section{Backgrounds}

The main backgrounds are expected to arise from Drell-Yan production in association with 
four or more jets, dilepton top-quark events, and QCD multijet events. The latter is 
the dominant background for the electron 
channel (followed by the Drell-Yan background). In the case of muons, the background is dominated by the Drell-Yan and top pair 
production. We used Monte Carlo to calculate background from the first two sources, and data to estimate background from 
QCD jets.

Background for the Run 1 analysis was estimated to be $1.8\pm 0.2\pm 0.3$ (with $1.27 \pm 0.24$ from QCD and 
$0.42\pm 0.15\pm 0.16$ from the other processes) for $\sim 100\ pb^{-1}$ of data. To extrapolate this number to 
the data set from Run 2, we 
have simply multiplied it by the ratio of luminosities to obtain $36\pm 4 \pm 6$ events. However, it is expected that due to the 
central magnetic field in the upgraded D\O\ detector, the probability of jets to be misidentified as electrons will be reduced by 
a factor of $\sim 2$ in Run 2. We have therefore considered a second scenario with the smaller expected background of 
$15 \pm 1.5 \pm 1.5$ events.

For the muon channel, the expected background has been scaled directly  from the Run 1 analysis. We expect $10 \pm 1 \pm 1$ 
background events in Run 2.

\section{Results}

In order to obtain the sensitivity of Run 2 in to RPV decays, we calculated the efficiency for signal for 
all the mass points shown in Fig.~\ref{fig:RPV-points}. Typical efficiencies,
the signal cross section in the $ee + 4$~jets 
channel, and the 
expected event yield in 2~$fb^{-1}$ of data, for several representative $(m_0,m_{1/2})$ points, are given in 
Table~\ref{table:RPV-eff}. Similar numbers are obtained for the muon channel.

\begin{table}[htb!]
\begin{center}
\caption{Efficiency $\times$ BR (\%), signal cross section and the expected event 
yield in 2~$fb^{-1}$ of data, at various $(m_0,m_{1/2})$ parameter space points. }
\begin{tabular}{||c|c|c|c|c||}
\hline
$m_0$ & $m_{1/2}$ & Efficiency $\times$ BR (\%)
      &  Cross section &  $\langle N \rangle$        \\
 (GeV) & (GeV) & & (pb) & (in 2 $fb^{-1}$) 	     \\
\hline
60    & 235 & $7.9\pm 1.1 $ & 0.16 & $25.2 \pm 3.4 $ \\
60    & 245 & $8.3\pm 1.1 $ & 0.08 & $12.8 \pm 1.7 $ \\
60    & 255 & $8.3\pm 1.1 $ & 0.06 & $10.5 \pm 1.4 $ \\
100   & 220 & $6.1\pm 0.8 $ & 0.10 & $12.2 \pm 1.7 $ \\
100   & 230 & $7.0\pm 1.0 $ & 0.08 & $11.3 \pm 1.5 $ \\
180   & 240 & $7.0\pm 0.9 $ & 0.05 & $7.1  \pm 1.0 $ \\
320   & 240 & $7.1\pm 0.9 $ & 0.05 & $6.9  \pm 1.0 $ \\
\end{tabular}
\end{center}
\label{table:RPV-eff}
\end{table}

\begin{figure}[t]
\centerline{\protect\psfig{figure=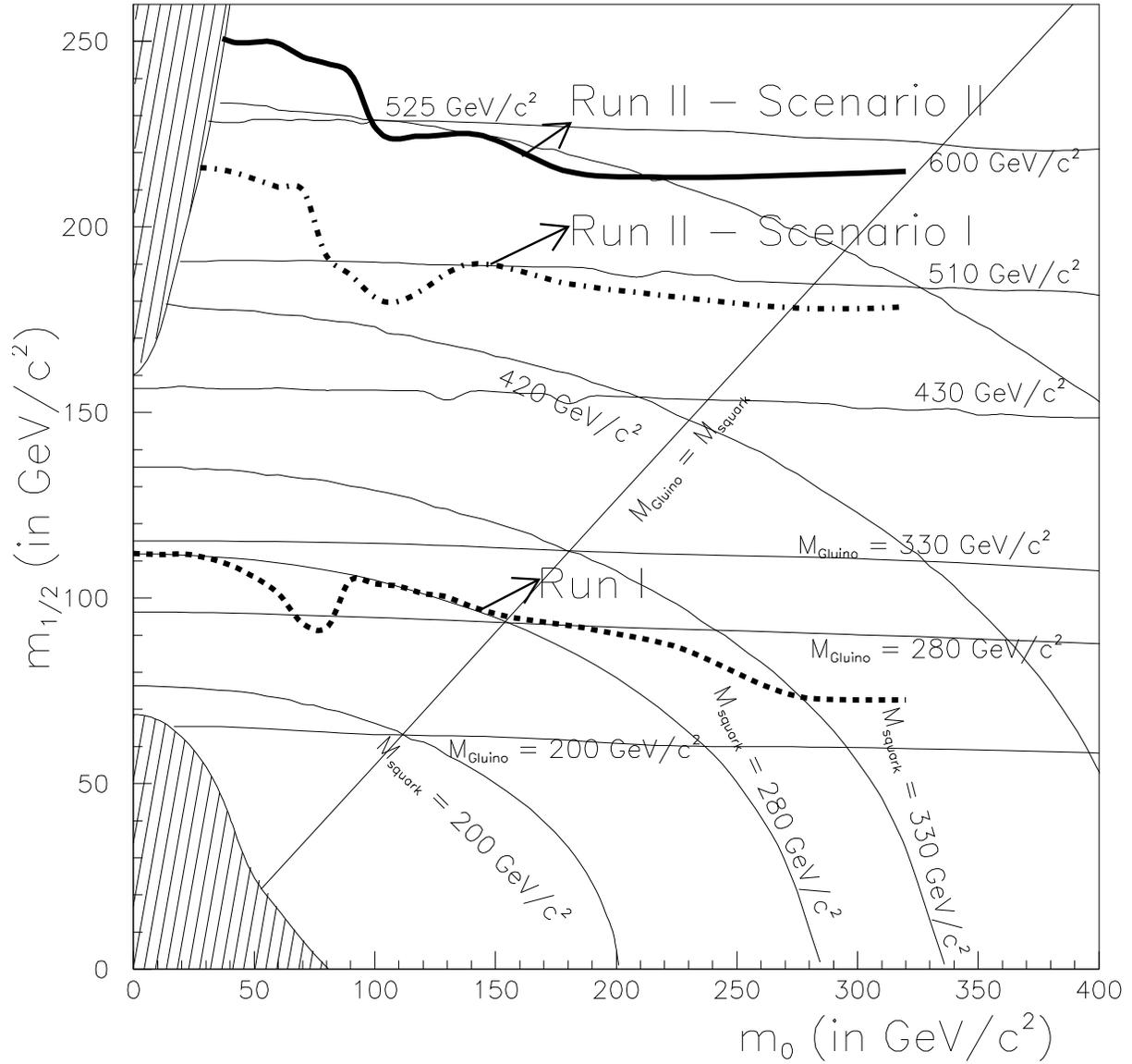,width=\textwidth}}
\caption{Estimated exclusion contour for Run 2 in the $(m_0,m_{1/2})$ plane for 
$tan \beta = 2$, $A_0=0$, $\mu <0$, from the $ee + 4$~jets channel.
Scenario I corresponds to a background of $36\pm 4 \pm 6$ events (direct scaling from Run 1); scenario II uses the background 
of $15 \pm 1.5 \pm 1.5$ events (scaling, but with improvements in the detector taken into account).}
\label{fig:RPV-e}
\end{figure}

\begin{figure}[t]
\vspace*{0.1in}
\centerline{\protect\psfig{figure=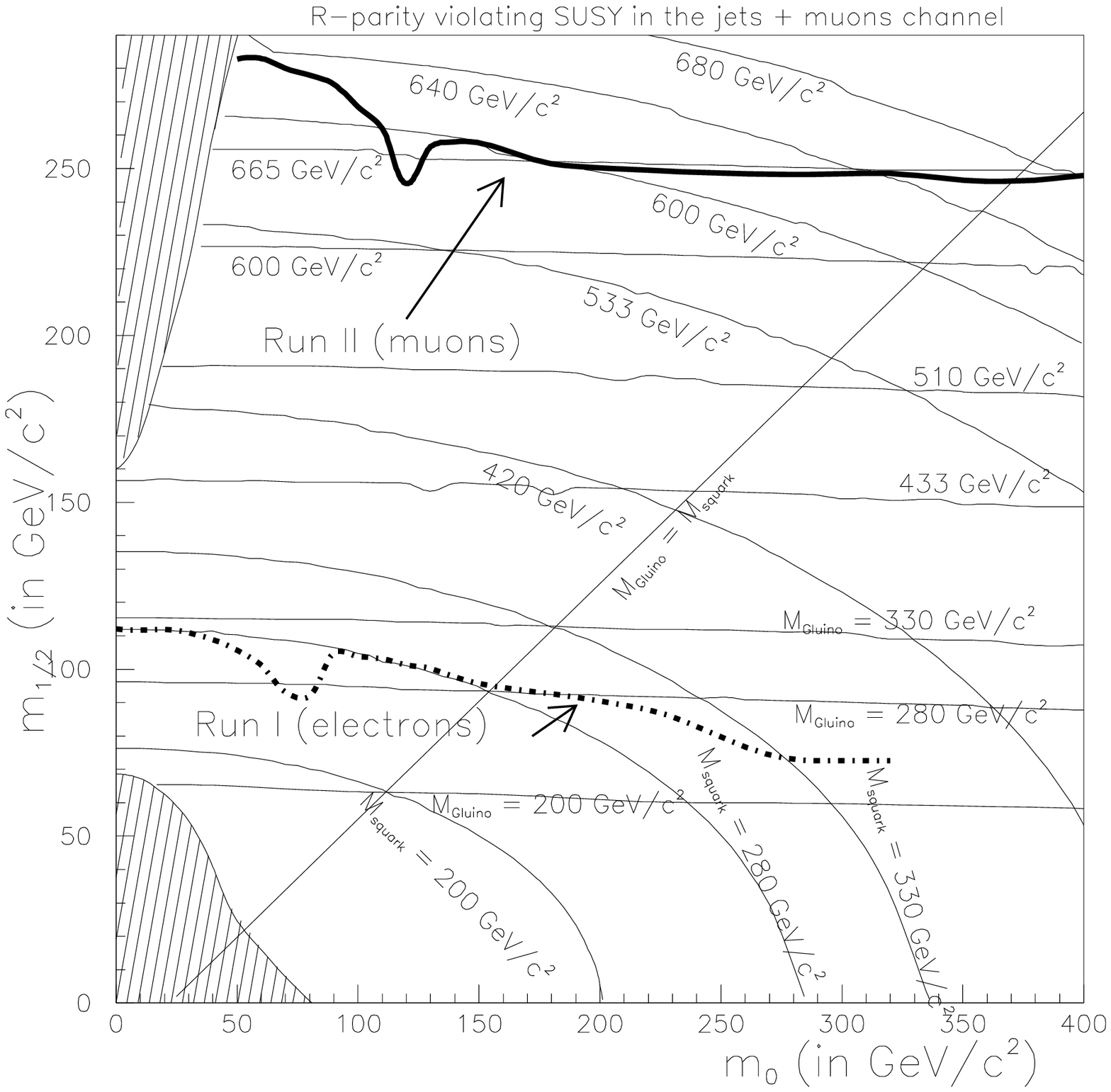,width=\textwidth}}
\caption{Estimated exclusion contour for Run 2  in the $(m_0,m_{1/2})$ plane for 
$tan \beta = 2$, $A_0=0$, $\mu <0$, from the $\mu\mu + 4$~jets channel for
background of $10 \pm 1.0 \pm 1.0$ (direct scaling from Run 1).}
\label{fig:RPV-mu}
\vfill
\end{figure}

We use these efficiencies to obtain exclusion limits in the $(m_0,m_{1/2})$ plane at 95\% CL, assuming that no excess of 
events will be observed above the predicted background. The exclusion contours for the electron and muon channel are shown in 
Fig.~\ref{fig:RPV-e} and \ref{fig:RPV-mu}, respectively. Numerical values of the limits are summarized in 
Table~\ref{table:RPV-limits}.

\begin{table}[hbt!]
\begin{center}
\caption{Lower limits on the squark and gluino masses from Run 2.}
\begin{tabular}{||c|c|c|c||}
\hline
~ & Lower limit on $m_{\tilde{q}}$ & Lower limit on $m_{\tilde{g}}$ & Limit when
                                       $m_{\tilde{q}} = m_{\tilde{g}}$ \\
~ & ( For any $m_{\tilde{g}}$)  & ( For any $m_{\tilde{q}}$ ) & \\\hline
\multicolumn{4}{||c||}{Electrons}\\
\hline
Run 1 & 252 GeV & 232 GeV & 283 GeV \\
Run 2 (Scenario I) & 430 GeV & 490 GeV & 490 GeV \\
Run 2 (Scenario II) & 520 GeV & 575 GeV & 585 GeV \\\hline
\multicolumn{4}{||c||}{Muons}\\\hline
Run 2 & 560 GeV & 640 GeV & 665 GeV \\
\end{tabular}
\end{center}
\label{table:RPV-limits}
\end{table}

It's worth mentioning that our analysis provides a conservative estimate of the sensitivity achievable in Run~2, since no formal 
optimization of the signal vs. background has been performed. We expect that a formal optimization can improve the sensitivity 
in the mass reach by 15--20\%.

\vfill

\end{document}